\documentclass[%
 aip,
 sd,%
 amsmath,amssymb,
reprint,%
author-year,%
]{revtex4-1}

\usepackage{graphicx}
\usepackage{dcolumn}
\usepackage{bm}
\usepackage{subcaption}
\usepackage{color,paralist}
\graphicspath{{./figures/}}
\definecolor{orange}{RGB}{255,127,0}
\begin{document}

\newcommand{\eq}{\mathrm{eq}}
\newcommand{\e}{\mathrm{e}}
\renewcommand{\d}{\mathrm{d}}
\newcommand{\ree}{$r_{\mathrm{ee}}$}
\newcommand{\res}{$r_{\mathrm{es}}$}
\newcommand{\re}{$r_{\mathrm{e}}$}
\newcommand{\rel}{$r_{\mathrm{el}}$}
\newcommand{\rl}{$r_{\mathrm{l}}$}

\newcommand{\replyc}[1]{{\color{blue}#1}}

\preprint{AIP/123-QED}

\title[The MPT as a result of bistability]{The Mid-Pleistocene Transition induced by delayed feedback and bistability}

\author{Courtney Quinn}
 \altaffiliation{c.quinn2@exeter.ac.uk}
\author{Jan Sieber}%
\affiliation{ 
College of Engineering, Mathematics and Physical Sciences, \\ University of Exeter, Exeter EX4 4QE, United Kingdom
}%

\author{Anna S. von der Heydt}
\affiliation{%
Institute for Marine and Atmospheric Research,
Centre for Extreme Matter and Emergent Phenomena,
Utrecht University, Princetonplein 5, 3584 CC
Utrecht, The Netherlands
}%

\author{Timothy M. Lenton}
\affiliation{%
Earth System Science, College of Life and Environmental Sciences, \\ University of Exeter, Exeter EX4 4QE, United Kingdom
}%

\date{\today}

\begin{abstract}

The Mid-Pleistocene Transition, the shift from 41 kyr to ~100 kyr glacial-interglacial cycles that occurred roughly 1 Myr ago, is often considered as a change in internal climate dynamics.  Here we revisit the model of Quaternary climate dynamics that was proposed by Saltzman and Maasch (`Carbon cycle instability as a cause of the late Pleistocene ice age oscillations: modelling the asymmetric response'. Glob Biogeochem Cycle 1988; 2: 177-185-from this point, referred to as SM88).  We show that it is quantitatively similar to a scalar equation for the ice dynamics only when combining the remaining components into a single delayed feedback term.  The delay is the sum of the internal time scales of ocean transport and ice sheet dynamics, which is on the order of 10 kyr.  

We find that, in the absence of astronomical forcing, the delayed feedback leads to bistable behaviour, where stable large-amplitude oscillations of ice volume and an equilibrium coexist over a large range of values for the delay.  We then apply astronomical forcing using the forcing data for integrated summer insolation at 65 degrees North from Huybers and Eisenman (\textit{Integrated Summer Insolation Calculations.} NOAA/NCDC Paleoclimatology Program Data Contribution \# 2006-079, 2006).  Since the precise scaling of the forcing amplitude is not known, we perform a systematic study to show how the system response depends on the forcing amplitude. We find that over a wide range of forcing amplitudes the forcing leads to a switch from small-scale oscillations of 41 kyr to large-amplitude oscillations of roughly 100 kyr without any change of other parameters.  The transition in the forced model consistently occurs at about the same time as the Mid-Pleistocene Transition (between 1200 and 800 kyr BP) as observed in the data records from Lisiecki and Raymo (`A Pliocene-Pleistocene stack of 57 globally distributed benthic $\delta^{18}$O records'. \textit{Paleoceanography} 2005; 20:1-17)..  This provides evidence that the MPT could have been primarily forcing-invoked.  Small additional random disturbances make the forcing-induced transition near 800 kyr BP even more robust.

We also find that the forced system forgets its initial history during the small-scale oscillations, in particular, nearby initial conditions converge prior to transitioning.  In contrast to this, in the regime of large-amplitude oscillations, the oscillation phase is very sensitive to random perturbations, which has a strong effect on the timing of the deglaciation events.

\end{abstract}

\maketitle

\section{\label{sec:intro}Introduction}
\begin{figure}
\includegraphics[width=0.5\textwidth]{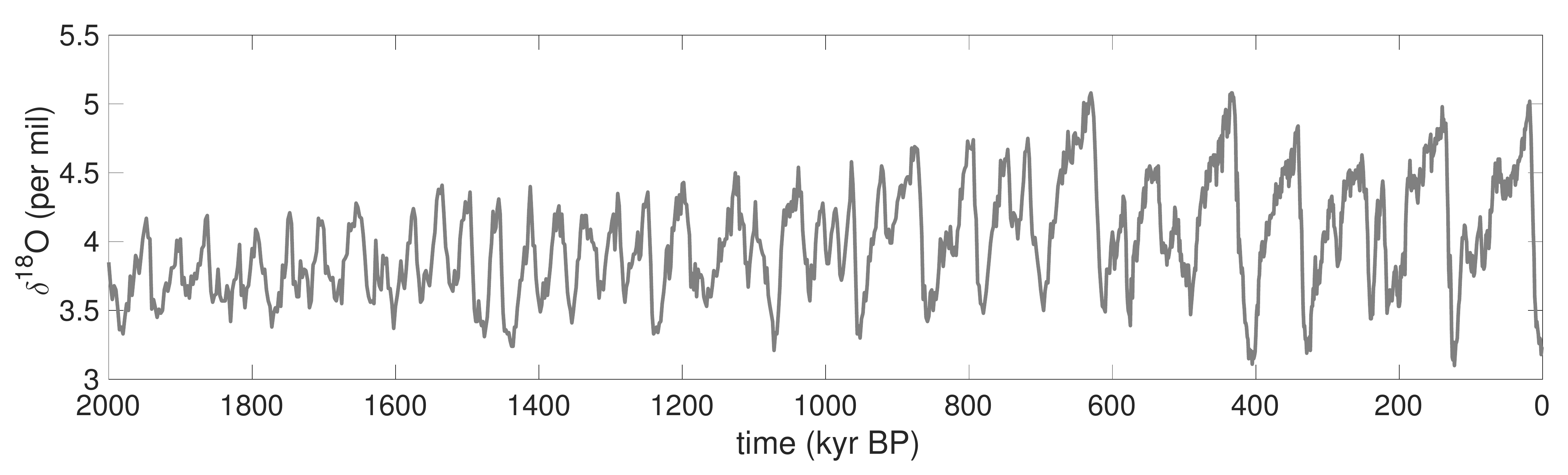}
\caption{\label{fig:record} Compilation of 57 globally distributed benthic $\delta^{18}$O records for the last 2 Myr \citep{lisiecki2005}.}
\end{figure}
The Pleistocene is characterised by its successive glacial-interglacial cycles whose documented periodicities has been a subject of interest for decades.  \citet{hays1976,paillard1998}, and \citet{ganopolski2011} have all presented evidence that orbital forcing plays a large role in the onset and termination of glacial periods, but the full extent of the variations cannot be explained solely by forcing \citep{ridgwell1999,shackleton2000,paillard2001}.  One actively researched aspect of the Pleistocene is the abrupt change in frequency of major glaciations known as the Mid-Pleistocene Transition (MPT) \citep{pisias1981,maasch1988,maasch1990,paillard1998,clark2006,crucifix2012,ashwin2015}.  Fig. \ref{fig:record} shows a proxy record compiled by \citet{lisiecki2005} for global temperature taken from ocean core sediments which can be directly related to global ice cover.  Spectral analysis has been performed on this time series, and it has been observed that the signal of the ~100-kyr cycle began to rise 1250 kyr BP and was established as the dominant cycle by 700 kyr BP \citep{clark2006,lisiecki2007,dijkstra2013}.  Many researchers have tried to determine not only the cause of this switch, but also the driving force behind the 100 kyr cycles themselves \citep{maasch1990,saltzman1991,paillard1998,gildor2001,paillard2004,ganopolski2011,ashwin2015}.  While the 41-kyr oscillations have been attributed to external forcing \citep{paillard2001}, the 100-kyr cycles have been proposed to be a result of nonlinear responses of the climate system itself \citep{imbrie1993,yiou1994,gildor2001}.

The major glacial-interglacial cycles were originally attributted to changes in insolation due to orbital variations (i.e. the Milankovitch cycles).  Milankovitch argued that summer insolation at high northern latitudes determined the main pacing of glaciations, as these changes control the length of seasons and amount of solar energy being recieved in high latitudes where the ice resides.  The most prominent frequencies of these changes are 19 and 23 kyr due to precession and 41 kyr due to obliquity \citep{paillard2001}.  Precession is the orientation of the rotational axis of the earth (axial) and the rotation of the orbital axis (apsidal), and obliquity is the angle between the rotational axis and the orbital axis.  \citet{milankovitch1941} argued that the main driving forces of climate fluctuations are obliquity and precession, which agrees with the records prior to the mid-Pleistocene transition.  There is a smaller signal of orbital forcing that correlates with the 100-kyr oscillations \citep{imbrie1993}.  Eccentricity, the amount the earth's orbital ellipse deviates from a circle, shifts with a main period of 413 kyr, but there are components that vary with periods between 95 kyr and 125 kyr.  The amplitude of these forcing signal bands, however, are an order of magnitude smaller than the signals of the 23- and 41-kyr bands \citep{hays1976}.
\\ \hspace*{1em} Since Milankovitch's theory, there have been many other attempts to identify the cause of the MPT (see review paper by \citet{crucifix2012}).  The starting point of our analysis is the collection of Saltzman models from the late twentieth century.  They were a collection of attempts to model the transition as a bifurcation in low-order dynamical systems due to climate feedbacks.  The models typically only had three dynamic variables usually representing some measure of global ice volume, global mean temperature, and ocean circulation strength. Other modelling attempts, after the Saltzman models, include (in order of increasing complexity) conceptual models \citep{paillard2017}, threshold models \citep{tzedakis2017}, nonsmooth models \citep{paillard1998}, relaxation oscillators \citep{ashwin2015}, box models of intermediate complexity \citep{gildor2001,tziperman2003}, and Earth system models of intermediate complexity (EMICs) \citep{ganopolski2011}.  While all of these studies were able to reproduce a transition similar to the MPT, they are all based on the hypothesis that the MPT occured because of some significant change in the dynamics of the system.  \citet{huybers2009} proposed an alternative hypothesis that the MPT was a spontaneous response to the astronomical forcing.  In this study we focus on this alternative hypothesis, and provide more evidence using a previously established model by \citet{saltzman1988}.
\\ \hspace*{1em} The paper is arranged as follows.  Section 2 revisits and analyses the three-variable Saltzman and Maasch 1988 model.  We observe that two of the variables mostly act as delays in feedback loops such that we may reduce the model to a scalar equation with delayed feedback.  We also identify a region of bistability at the parameter values of interest which was not explored in the original model analysis.  Section 3 shows the model response in the bistable region when astronomical forcing is introduced, which includes MPT-like realisations with no change in parameters.  In section 4 we investigate the sensitivity of the model. We observe that the system forgets its initial condition and has low sensitivity whenever it exhibits a small-amplitude oscillatory response.  It is much more sensitive to disturbances when exhibiting a large-amplitude oscillatory response.  This sensitivity affects mostly the timing of major deglaciation events.  Section 5 discusses our results.

\section{Adaptation of the Saltzman and Maasch model}

We begin by considering the original model of \citet{saltzman1988}, which we will refer to as SM88, 
\begin{subequations} \label{eq:MS88}
\begin{align}
\dot{X}(t) &= -X(t)-Y(t), \label{eq:MS88-X} \\
\dot{Y}(t) &= -pZ(t)+rY(t)+sZ(t)^{2}-Z(t)^{2}Y(t), \label{eq:MS88-Y} \\
\dot{Z}(t) &= q(-X(t)-Z(t)). \label{eq:MS88-Z}
\end{align}
\end{subequations}
Here $X,Y,Z$ are scaled versions of global ice mass, atmospheric CO$_{2}$, and North Atlantic Deep Water (NADW) respectively.  More precisely, the variables represent anomalies from a background state.  The first term in (\ref{eq:MS88-X}) represents the feedbacks of global ice mass.  The combined effect of damping and negative feedback from NADW production is taken to outweigh the positive feedback from ablation (melting and loss of ice through icebergs).  The second term in (\ref{eq:MS88-X}) is the direct effect of higher temperatures (caused by higher CO$_2$ levels) leading to loss of ice mass.  In equation (\ref{eq:MS88-Z}) the negative sign of ice anomaly $X$ is motivated by the reduction of NADW production with loss of ice mass.  This equation has a time scaling $q$ which is the ratio of ice sheet time constant (10 kyr, which equals $t=1$ in the scaling of (\ref{eq:MS88-X})-(\ref{eq:MS88-Z})) to the response time of the deep ocean.  The equation for atmospheric CO$_{2}$ (\ref{eq:MS88-Y}) has a negative term $-pZ$ for the negative effect due to NADW- more NADW leads to more ventilation of the deep ocean (stronger mixing between surface and deep waters) and an increased downdraw of atmospheric CO$_{2}$ into the deep ocean.  The term $rY$ accounts for the positive feedbacks of atmospheric CO$_{2}$, related to changes in sea surface temperatures, sea ice extent, and sea level, which outweigh negative feedbacks.  The nonlinearity $(s-Y)Z^2$ is motivated by the appearance of locally enhanced instabilities in the Southern Ocean due to increased NADW production.  The NADW meets colder, denser Antarctic Bottom Water and induces vertical mixing.  This brings CO$_2$-rich water to the surface which releases CO$_2$ into the atmosphere depending on the current level of atmospheric CO$_2$ ($Y$).  Each of these effects (except damping of the nonlinear terms) have an associated parameter which controls their relative strengths.  These were the feedbacks between global ice mass, atmospheric CO$_{2}$, and NADW deemed most important by \citet{saltzman1988}.  Present day studies continue to stress the interaction between these three variables as being essential for the MPT, including a study by \citet{chalk2017} which uses a geochemical box model to confirm that carbon cycle feedbacks related to ocean circulation are necessary to sustain the long glacial cycles in the late Pleistoncene.

A change of variable $V(t) = -X(t)$, where $V$ is the negative of the global ice mass perturbations, replaces the two linear equations (\ref{eq:MS88-X}) for $X$ and (\ref{eq:MS88-Z}) for $Z$ by a chain of linear first-order filters
\begin{align} \label{eq:MS88-LCA}
\dot{V}(t) &= Y(t)-V(t), &
\dot{Z}(t) &= q [V(t)-Z(t)].
\end{align}
A linear chain of filters is well known to approximate a delay \citep{smith2010}:\begin{equation}\label{eq:q2tau}
Z(t)\approx Y(t-\tau)\mbox{, where\quad  }\tau=1+\frac{1}{q}\mbox{.}
\end{equation}
Appendix~\ref{app:LCA} gives further comments on the linear chain
approximation for delays.  Using this connection between linear chains and delays, we can rewrite the model as a scalar delay-differential equation (DDE) for $Y$.  To facilitate the comparison to data at a later stage, we shift time and exploit that in the DDE, the ice mass anomaly $V=-X$ is just the delayed (by time 1) CO$_2$ anomaly: $X(t)=-Y(t-1)$. 
Thus, we may express the DDE
  model in terms of ice mass anomaly $X$:
\begin{equation} \label{eq:DDE}
\dot{X}(t) = rX(t)-pX(t-\tau) 
-X(t-\tau)^{2}[s+X(t)]\mbox{.}
\end{equation}
In DDE \eqref{eq:DDE} one of the parameters $r$, $p$, $s$, $\tau$
is redundant and could be removed by a rescaling of $X$ and
time. Thus, \eqref{eq:DDE} is objectively simpler than the original
SM88 model \eqref{eq:MS88} as it has fewer free parameters. For the purpose of model comparison, the DDE system will be left as above.

\subsection{Dynamics of DDE, compared to SM88}

We initially perform a systematic analysis of DDE~\eqref{eq:DDE}
  and compare it to the original SM88.  We find that models (\ref{eq:MS88}) and (\ref{eq:DDE}) have the same long-time behaviour (using the relation $q = \frac{1}{\tau-1}$ implied by \eqref{eq:q2tau}).

Since delays, whether in the form of chains as in SM88 or explicit, don't affect location of equilibria, i.e. $X(t)=X(t-\tau)$ (or $X(t)=-Y(t)=-Z(t)$), equilibria for both models, \eqref{eq:MS88} and \eqref{eq:DDE}, (called $X_{\eq,j}$) satisfy
\begin{equation}
0 = -pX_{\eq}+rX_{\eq}- sX_{\eq}^{2}-X_{\eq}^{3}. \label{equilibria}
\end{equation}
Thus, we have a trivial equilibrium $X_{\eq,1}=0$ for all parameter values and non-zero equilibria $X_{\eq,2,3}$ for $s^{2}>4(p-r)$:
\begin{equation} \label{eq:equilibria}
X_{\eq_{2,3}} = \frac{1}{2}\left[-s \pm \sqrt{s^{2}-4(p-r)}\right]\mbox{.}
\end{equation}
The stability of the equilibria may differ, however, since eigenvalues of the linearisation may cross the imaginary axis (leading to oscillations in a Hopf bifurcation \citep{kuznetsov2013}) at different parameter values for (\ref{eq:MS88}) and (\ref{eq:DDE}).

All other trajectories have to be studied numerically (for both,
  SM88 \eqref{eq:MS88} and DDE\eqref{eq:DDE}). Since the scaling between the two models are the exactly same, we will
use values for $p$, $r$, and $s$ from \citet{saltzman1988} for all numerical studies:
\begin{align}
  p&=0.95\mbox{,}& r&=0.8\mbox{,} & s&=0.8\mbox{,}\label{eq:pars}
\end{align} 
while varying the timescale (or delay) of the feedback
processes ($\tau$ for DDE~\eqref{eq:DDE}, $q$ for SM88 model
\eqref{eq:MS88}). A value of $q=1.2$ was used in SM88, which would correspond to $\tau\approx1.83$.  The only constraint on $q$ was $q\geq 1$ ($q$ was defined as the ratio of ice sheet time constant, 10 kyr, to the slow response time of deep ocean).  This would put realistic values of $\tau$ in the range [1,2].  The value of $q$ by \citet{saltzman1988} would approximate a deep ocean response time of around 8.3 kyr.  This is long for ocean timescales as is shown through simulations of general circulation models.  For example, \citet{yang2011} studied the response of the deep ocean under varying climate scenarios and found it to be 1.5 to 2 kyr.  While arguments for a slightly longer ice sheet response or ocean response could be made, $\tau<1.6$ is more realistic that the suggested $\tau=1.83$ in SM88. The comments below \eqref{eq:DDE} also show that the change  $\tau\to\tau/\alpha$, $p\to\alpha p$, $r\to\alpha r$, $s\to\sqrt{\alpha}s$ leaves the system unchanged such that all phenomena reported in this paper are observable for any delay and appropriately rescaled parameters.

\begin{figure*}
\begin{subfigure}{\textwidth}
	\includegraphics[width=0.8\textwidth]{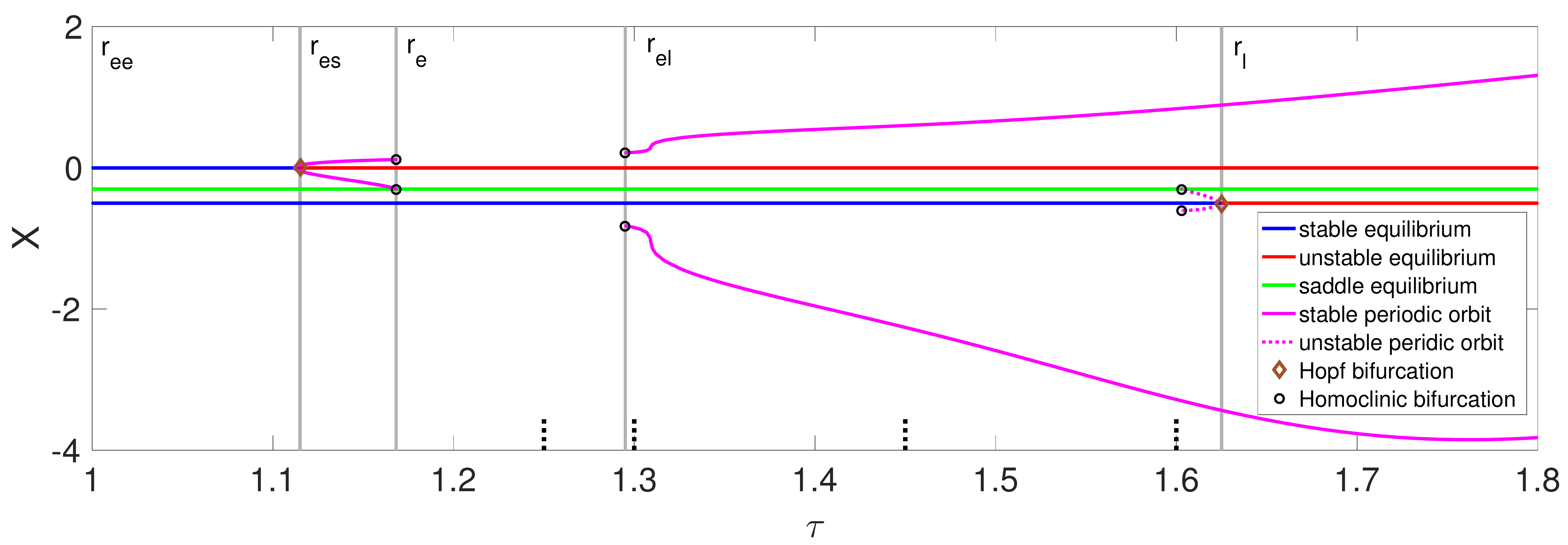}
	\caption{\label{fig:bifdiag_dde} DDE model (\ref{eq:DDE})  \\The dotted black lines
	indicate values of $\tau$ used in section \ref{sec:MPT}: $\tau_{\mathrm{ref}}=1.25$,
	 $\tau_1=1.3$, $\tau_2=1.45$, $\tau_3=1.6$.}
\end{subfigure}
\begin{subfigure}{\textwidth}
\includegraphics[width=0.8\textwidth]{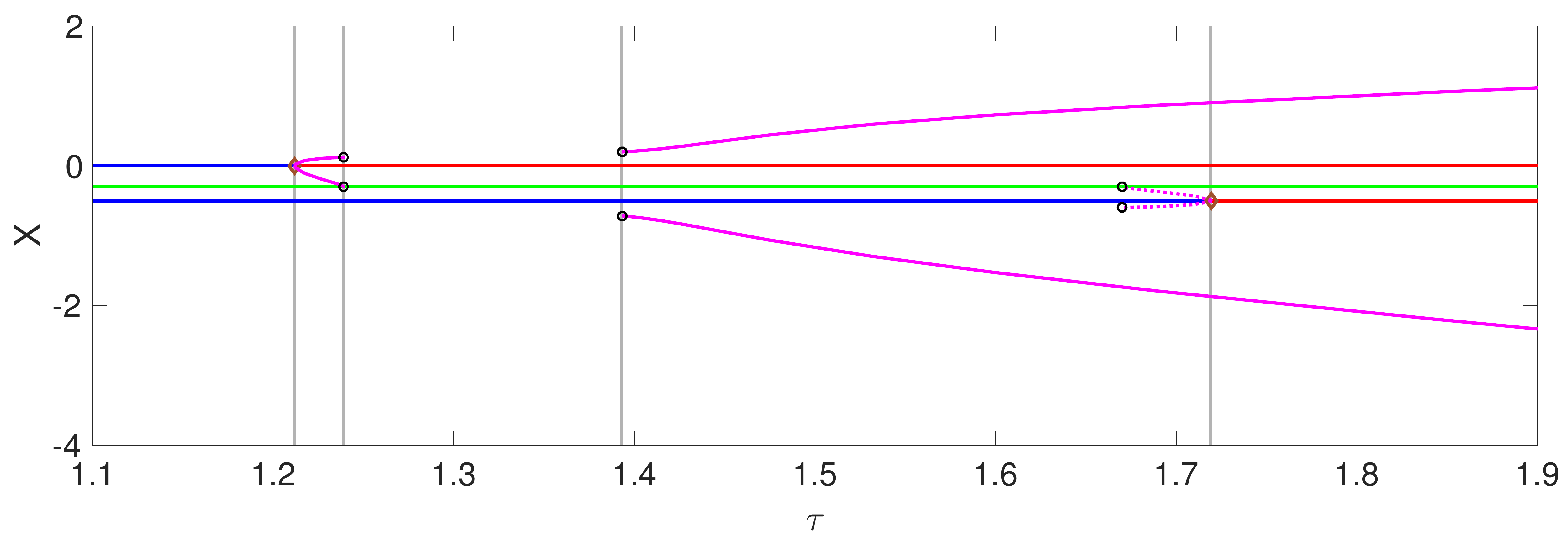}
	\caption{\label{fig:bifdiag_orig}SM88 ($\tau = \frac{1}{q}+1$)}
\end{subfigure}
\caption{\label{fig:bifdiag} Bifurcation diagrams of both models for delay parameter $\tau$.
	Other parameters: $p=0.95$, $r=0.8$, $s=0.8$.}
\end{figure*}

Figure \ref{fig:bifdiag} shows the bifurcation diagrams for varying $\tau$ in DDE \eqref{eq:DDE} and
  SM88 model \eqref{eq:MS88}, respectively.  Although the locations of bifurcations are slightly shifted, qualitatively the two figures agree nicely.  In particular, even though the space of possible initial conditions for DDE~\eqref{eq:DDE} is infinite-dimensional (every
possible history on $[-\tau,0]$ gives a different trajectory), the
long-time behaviour of trajectories in the $(X(t),X(t-\tau))$-plane
follows a two-dimensional ordinary differential equation (ODE) in the range we explored.
The diagrams in Fig.~\ref{fig:bifdiag} are partitioned into five main regions of different global behaviour separated by grey vertical lines can be seen in both figures.  From left to right the attractors in each region are: 
\begin{compactitem}
\item {[\ree]} two stable equilibria,
\item {[\res]} one stable
  equilibrium and one stable small-amplitude periodic solution,
\item {[\re]} one stable equilibrium,
\item {[\rel]} one stable equilibrium
  and one stable large-amplitude periodic solution, and
\item {[\rl]} one
  stable large-amplitude periodic solution.
\end{compactitem}
The region of interest in this study is the bistable region with large-amplitude periodic orbits [\rel].  Note that [\rel] can be split further into two sections: one containing unstable periodic orbits and the other not.  The effect of the unstable periodic orbits is a drastically reduced basin of attraction for the stable equilibrium.  However, since there is no change in the attractors, we will consider these as one region.  Region [\rel] was not discussed in a later parameter study of the original model \citep{maasch1990}.  For the remainder of this paper we will discuss only the DDE model, but similar results can be seen for the ODE model.  

\subsection{The bistable region [\rel]}

\begin{figure}
\includegraphics[width=0.4\textwidth]{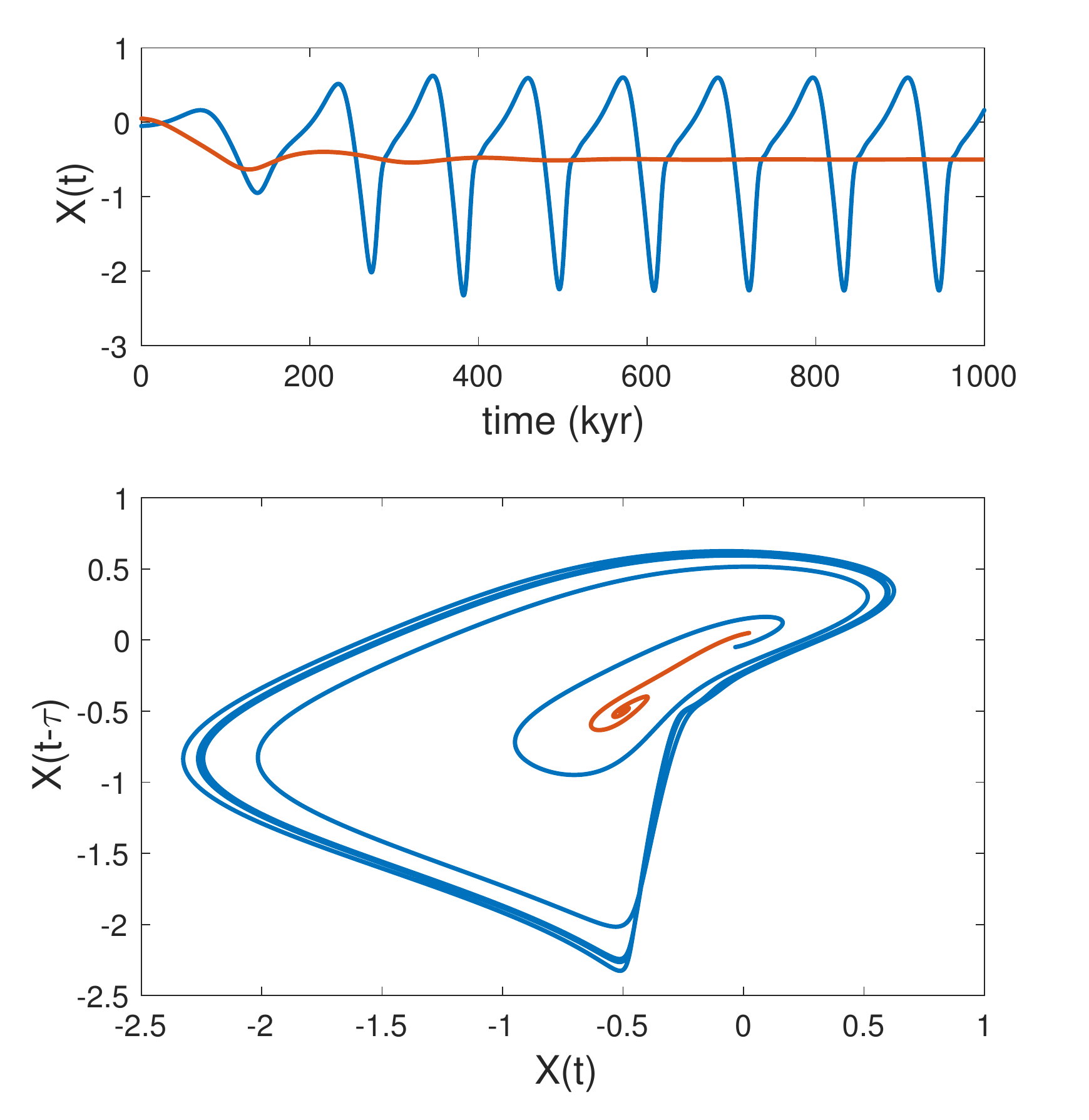}
\caption{\label{fig:bist_ex} Example trajectories in the bistable region [\rel] for $\tau=1.45$; $X(t) = -0.05$ (blue) and $X(t) = 0.05$ (red) for $t<0$. Top- time profiles, bottom - phase portraits.}
\end{figure}

The focus of this study is the bistability seen for $\tau\in[1.295,1.625]$ in the DDE system.  Not only is this a novel dynamical region, but it also fits within more realistic timescales of the ice sheet and deep ocean response. In this region there are two possible stable solutions (shown in Fig.~\ref{fig:bist_ex}): a stable equilibrium and stable large amplitude periodic orbits.  The time profile of the periodic orbits has the assymetrical shape observed in the ice age cycles of the late Pleistocene.  The assymetry is attributed to a slow accumulation of ice mass followed by rapid melting.  In addition, the period remains between 109 and 120 kyr throughout the bistable region (with an exception of $\tau$ very close to the [\re-\rel] boundary where the period approaches infinity).  This cycle length agrees with what is seen in the data (even more so when one adds the external forcing; see Sec.~\ref{sec:MPT}).  In previous studies of SM88 model \eqref{eq:MS88}, a transition between two stable states was enforced by a parameter shift through a Hopf bifurcation \citep{saltzman1988,maasch1990}.  We will demonstrate in the next section that the bistability in region [\rel] makes transitions
  between the two states possible without any change of parameter when the model is subjected to external forcing.

\section{The Mid-Pleistocene Transition under the forced model} \label{sec:MPT}

In this section we show that when adding astronomical forcing to model (\ref{eq:DDE}), a transition typically occurs at the same time as the MPT is seen in recorded data without further parameter tuning.  In section \ref{sec:forcing} we describe the astronomical forcing considered and how we include it in the model.  We then examine the responses for different delays and different forcing strengths in sections \ref{sec:varytau} and \ref{sec:varyu}.

\subsection{\label{sec:forcing}Astronomical forcing}

\begin{figure}
\includegraphics[width=0.5\textwidth]{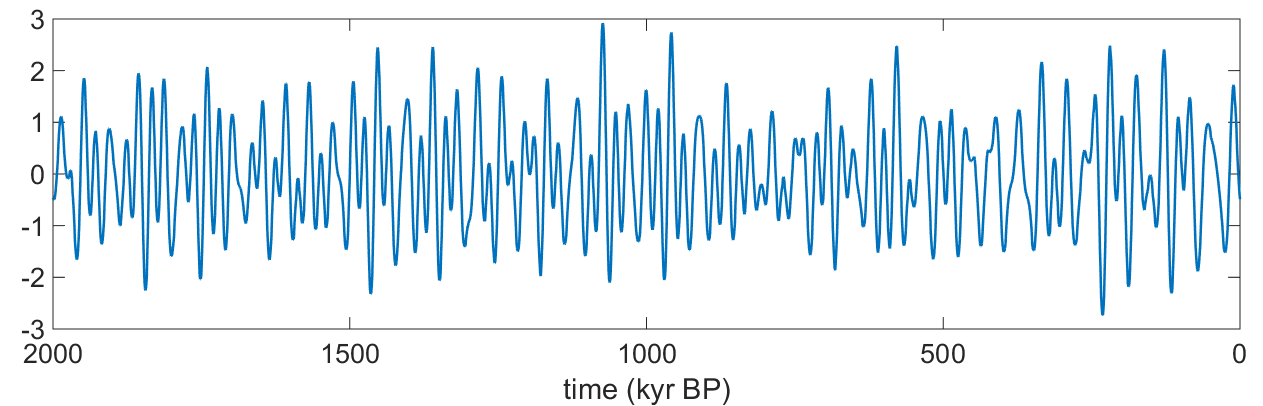}
\caption{\label{fig:forcing} Normalised integrated July insolation at 65$^{\circ}$N, adapted from \citet{huybers2006}.}
\end{figure}

\citet{hays1976} have provided evidence that the glacial cycles during the Pleistocene are driven primarily by variations in the earth's orbital cycle.  This includes changes in precession (orientation of the rotational axis), obliquity (angle between the rotational axis and orbital axis), and eccentricity (orbital ellipse's deviation from a circle).  These modes vary approximately periodically, with cycle lengths of 19/23 kyr, 41 kyr, and 100 kyr respectively \citep{milankovitch1941,berger1978, huybers2006}.  Fig.~\ref{fig:forcing} shows a time series of average daily summer insolation at 65$^{\circ}$N computed by \citet{huybers2006} based on the model introduced by \citet{huybers2006early}.
To investigate the effect of this forcing on DDE~\eqref{eq:DDE},
we include the astronomical forcing in the same way as in \citet{maasch1990}.  We add forcing signal
$M(t)$ shown in Fig.~\ref{fig:forcing} with negative
amplitude $u$,
\begin{equation} \label{eq:forced}
\dot{X}(t) = rX(t)-pX(t-\tau) 
-X(t-\tau)^{2}[s+X(t)] - uM(t)\mbox{.}
\end{equation}
The precise procedure for extracting $M(t)$ from the
  publicly available data source can be found in
  Appendix~\ref{app:forcing}.  We are interested in how the bistable
region responds to this external forcing in dependence of two
parameters: the delay $\tau$ and the forcing amplitude $u$.

For small $u$ the system is expected
to exhibit two types of responses to forcing in the bistable region.  Each of them is a perturbation of an attractor of the unforced system, namely the equilibrium and the large-amplitude periodic orbit, which persist for small $u$.  We will refer to these responses as the \emph{small-amplitude} and
the \emph{large-amplitude} response (compare red and blue time
profiles in Fig.~\ref{fig:exbist_tau16}). For increasing $u$ we
expect to observe increasingly frequent transitions between these
responses. For large forcing amplitudes $u$ the internal dynamics of
the model will be dominated by the forcing.

\subsection{\label{sec:varytau}Responses for different delays in the bistable region}

\begin{figure}
\centering
\begin{subfigure}[b]{0.5\textwidth}
        \includegraphics[width=\textwidth]{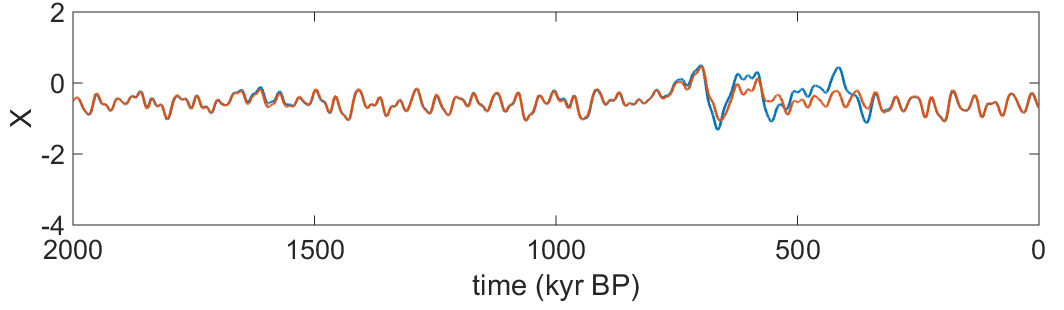}
        \caption{$\tau = \tau_1$ (blue), $\tau=\tau_{\mathrm{ref}}$ (red)}
        \label{fig:exbist_tau13}
\end{subfigure}
\begin{subfigure}[b]{0.5\textwidth}
        \includegraphics[width=\textwidth]{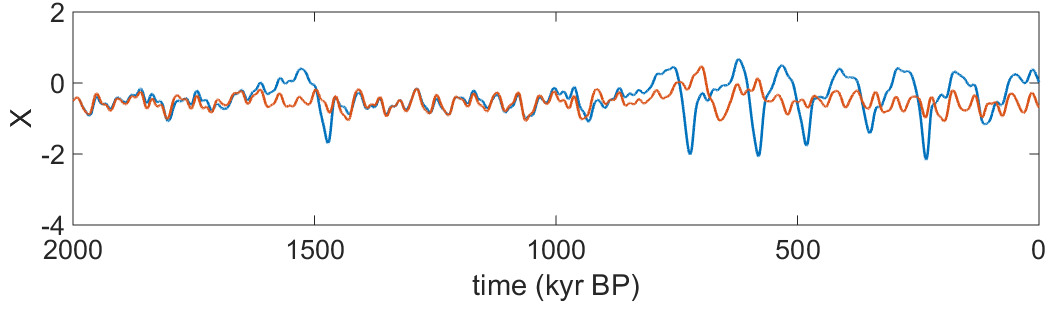}
        \caption{$\tau = \tau_2$ (blue), $\tau=\tau_{\mathrm{ref}}$ (red)}
        \label{fig:exbist_tau145}
\end{subfigure}
\begin{subfigure}[b]{0.5\textwidth}
        \includegraphics[width=\textwidth]{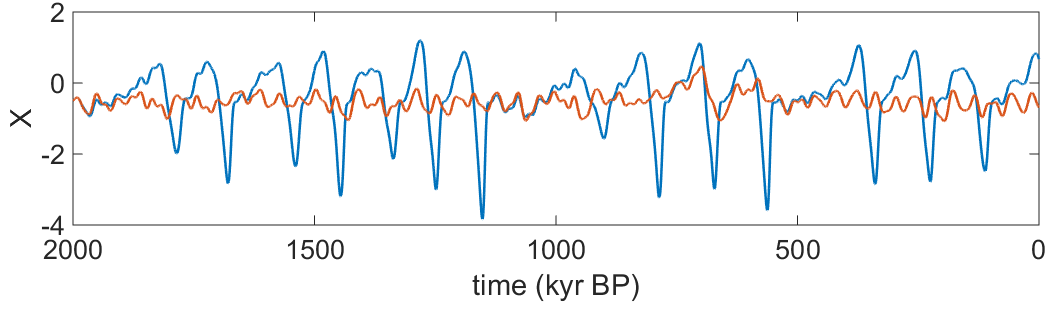}
        \caption{$\tau = \tau_3$ (blue), $\tau=\tau_{\mathrm{ref}}$ (red)}
        \label{fig:exbist_tau16}
\end{subfigure}
\caption{Trajectories in bistable region (blue) compared to a reference trajectory at $\tau_{\mathrm{ref}} = 1.25$ (red) for delays (a) $\tau_1 = 1.30$, close to homoclinic connection, (b) $\tau_2 = 1.45$, middle of bistable region, and (c) $\tau_3 = 1.60$, close to Hopf bifurcation.  Values of $\tau$ are indicated on bifurcation diagram in Fig.~\ref{fig:bifdiag_dde}.  Other parameters:
$p=0.95$, $r=0.8$, $s=0.8$, and $u=0.25$. Initial condition
$X(t)=-0.5$ for $t\in[2+\tau,2]$Myr BP in all cases.}
\label{fig:exbist}
\end{figure}

We choose a moderate value of the forcing amplitude $u=0.25$ and investigate the response at four values of $\tau$, labelled in Fig. \ref{fig:bifdiag_dde} by $\tau_{ref}$, $\tau_{1}$, $\tau_{2}$, and $\tau_{3}$.  The first value $\tau_{\mathrm{ref}} = 1.25$ is outside of the region of bistability and is used as a reference trajectory to which the solution trajectories for the other delays
$\tau_1$, $\tau_2$ and $\tau_3$ are compared.  The results are shown
in Fig.~\ref{fig:exbist} (response for $\tau_\mathrm{ref}$ shown in
red, for other delays in blue) 

The first comparison is made close to the boundary [\re]--[\rel] at $\tau_{1}=1.3$.  We see in figure \ref{fig:exbist_tau13} that both trajectories change in synchrony, exhibiting only the small-amplitude response.  In the middle of the bistable region
  ($\tau_{2}=1.45$, shown in Fig.~\ref{fig:exbist_tau145}), the solution shows a small-amplitude response in most of the first half of the time window and a large-amplitude response in the second half.  Close to the Hopf bifurcation of the autonomous stable equilibrium ($\tau_3=1.6$, shown in Fig.~\ref{fig:exbist_tau16}) the solution exhibits primarily a large-amplitude response.  This is expected due to the weakening attraction of the autonomous stable equilibrium and its shrinking basin of attraction. We will demonstrate in the following section that these transitions generate dynamics with time profiles that are qualitatively similar to the records of the MPT. 

\subsection{\label{sec:varyu}Variable forcing strength}

\begin{figure*}
\centering
\begin{subfigure}[b]{0.32\textwidth}
        \includegraphics[width=\textwidth]{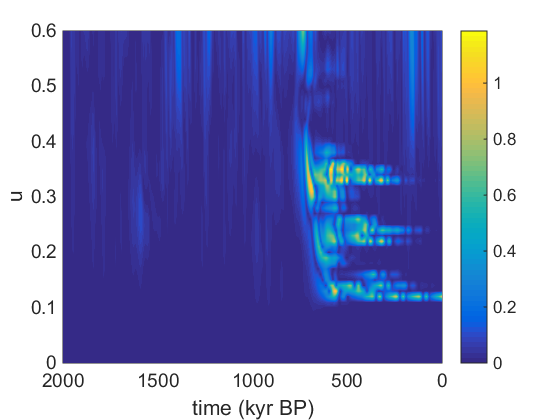}
        \caption{}
        \label{fig:disteq_tau13}
\end{subfigure}
\begin{subfigure}[b]{0.32\textwidth}
        \includegraphics[width=\textwidth]{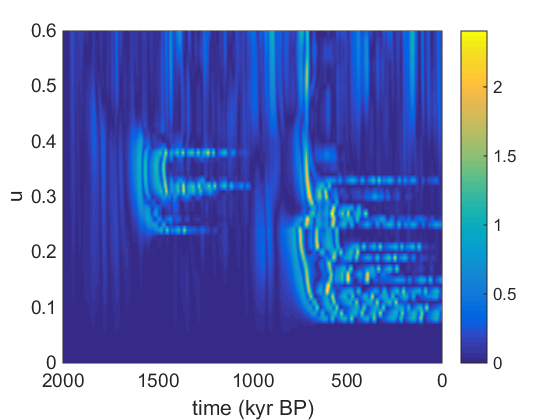}
        \caption{}
        \label{fig:disteq_tau145}
\end{subfigure}
\begin{subfigure}[b]{0.32\textwidth}
        \includegraphics[width=\textwidth]{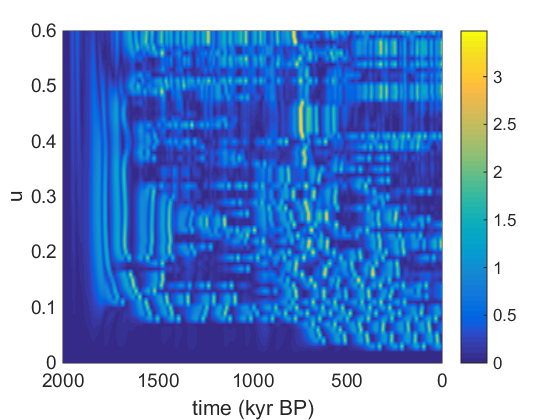}
        \caption{}
        \label{fig:disteq_tau16}
\end{subfigure}
\caption{Distance from reference solution at $\tau_{\mathrm{ref}}=1.25$, for different forcing strengths $u$.  Averages taken over window length of size $\tau$. (a) $\tau = 1.30$, close to homoclinic connection. (b) $\tau= 1.45$, middle of bistable region. (c) $\tau = 1.60$, close to Hopf bifurcation.  Initial condition $X(t)=-0.5$ for $t\in[2+\tau,2]$Myr BP} in all cases.
\label{fig:heatmaps}
\end{figure*}

For the exploration of the effect of different forcing strengths, we take the same values for $\tau$ as in Fig.~\ref{fig:exbist}, covering the range of the bistable region.  For each $\tau$ we compute trajectories for different $u$, ranging from $u=0$ to
$u=0.6$, with the initial history $X(t)=-0.5$ for $t\in[2+\tau,2]$Myr BP.  Figure~\ref{fig:heatmaps} shows the difference of
these trajectories to a respective reference trajectory for
$\tau_{\mathrm{ref}}=1.25$ and the same initial condition and forcing
strength $u$, averaged over a window of length $\tau$.  We note that differences to a reference trajectory at $u=0$ and identical delay $\tau$ (using same initial condition) give qualitatively similar results.

Remarkably, figures \ref{fig:disteq_tau13} and \ref{fig:disteq_tau145} show that there is a distinct period around 700-800 kyr BP where the solutions diverge from the reference trajectory in a large range of forcing strengths $u$.  This suggests that some aspect of the forcing around this time kicks the trajectories into the basin of attraction of the large-amplitude response.  An additional area like this is seen in figure \ref{fig:disteq_tau145} near 1600 kyr BP.  

The second interesting feature in all three figures is a threshold behaviour in the forcing strength $u$: below a certain value of $u$ specific to each $\tau$, transitions do not occur such that the solutions just track the reference trajectory.  Above this value trajectories suddenly can make this transition.  For some values of delay $\tau$ we observe an additional larger threshold value for the transition at 1600 kyr BP.

Thresholds in $u$ are also seen when periodic forcing is added to the model instead of the insolation time series.  We consider a new definition of the forcing function,
\begin{equation}
M(t) = \sin\bigg(\frac{2\pi}{4.1}t+\pi\bigg).
\end{equation}
This forcing has a period of 41 kyr.  If $u>0$, equation (\ref{eq:forced}) has two stable solutions: a sinusoidal signal with a 41 kyr period, or a large-amplitude asymmetrical quasiperiodic response. When recreating figure \ref{fig:disteq_tau145} with this periodic forcing, we see the same threshold transition at $u=0.08$.  We can show that this transition is due to moving basins of attraction for the two stable solutions.  The other sharp transition we saw in Fig.~\ref{fig:heatmaps} would not be possible with this forcing.  Either the solution would start to transition immediately, or it would not transition at all.  This is also the case if an envelope of modulated amplitude describing changes of eccentricity at $1.1$Myr periods is added for the periodic forcing.  This leads us to believe that the quasiperiodicity is necessary for a transition of this type.

\section{Model sensitivity to noise: desynchronisation and increased robustness of transition}

\subsection{Finite-time Lyapunov Exponents}

To analyse how trajectories depend on their history at specific instances of the forcing, we compute finite-time Lyapunov exponents (FTLEs) using a QR decomposition method (details in Appendix \ref{app:FTLE}).  This method was previous used in \citet{desaed2013} to illustrate the desynchronisation of nearby trajectories in a van der Pol-type oscillator model.  We will apply the ideas presentated in that study to our forced model.

The difference between FTLEs and classical Lyapunov exponents is that FTLEs are recorded for a family of time windows of finite length $w$ rather than over the entire long time run.  Thus, FTLEs are time-dependent functions instead of real numbers.  A positive FTLE along a given trajectory $X(t)$ at time $t_0$ indicates that some nearby trajectories diverge exponentially from $X(t)$ in the time window $[t_0-w,t_0]$.  Therefore, $X(t)$ is sensitive to small perturbations in the time window $[t_0-w,t_0]$.  While a trajectory could be asymptotically stable, a positive FTLE at a time $t_0$ indicates temporary amplification of perturbations from the attracting trajectory (observed as temporary desynchronisation, see \citep{desaed2013}).
\begin{figure}
\includegraphics[width=0.5\textwidth]{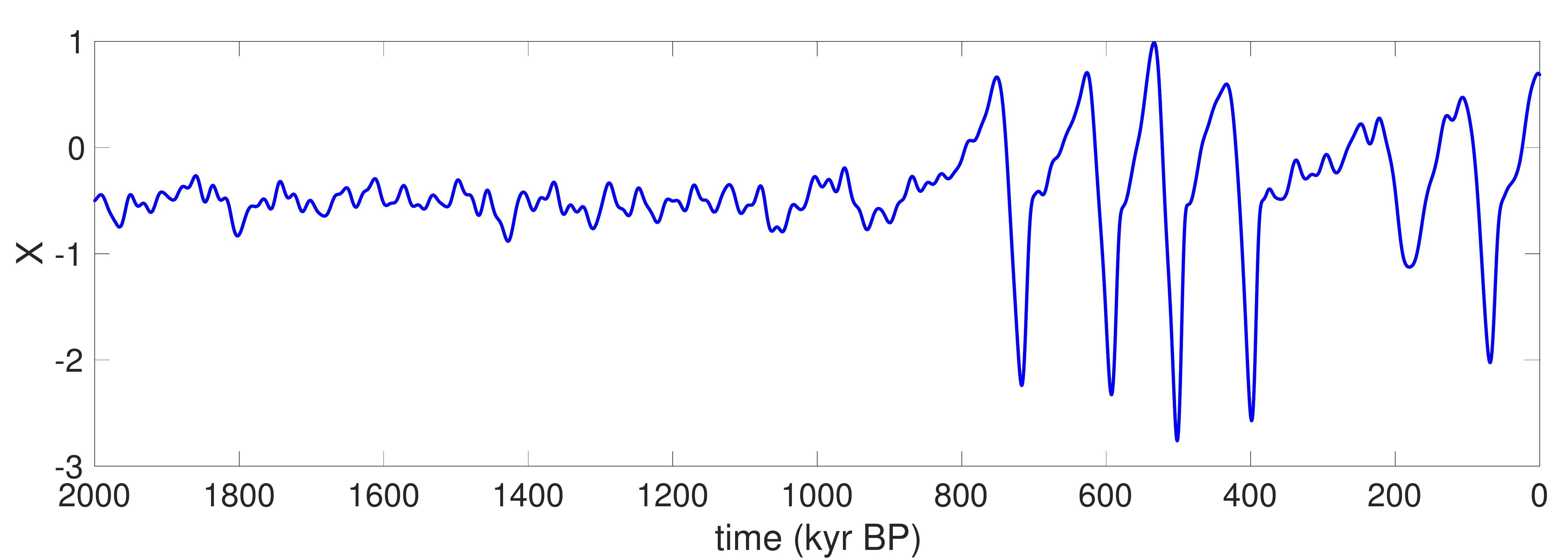}
\caption{Example trajectory that exhibits an MPT-like transition with $\tau = 1.45$ and $u = 0.15$.}
\label{fig:MPT}
\end{figure}

\subsection{FTLE implications on MPT and timing of major deglaciations}

We analyse one trajectory, showing a forcing-induced MPT ($\tau=1.45$, $u=0.15$,shown in Fig. \ref{fig:MPT})
.
We compute the FTLE over a sliding window of our example MPT tragectory with a window length of $w=250$ kyr.  This window length is chosen in order to filter out the dominant frequencies of the forcing.  Similar results are seen with any window lengths $w$ from 150-500 kyr.

\begin{figure}
\includegraphics[width=0.5\textwidth]{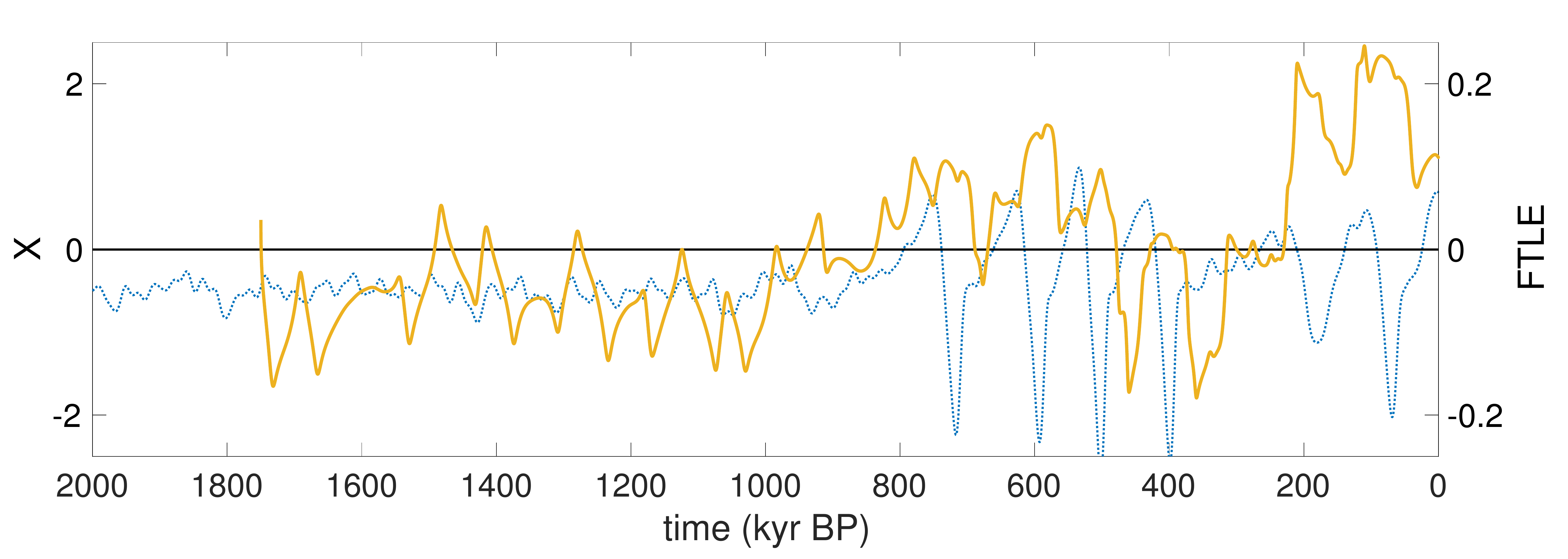}
\caption{Finite-time Lyapunov exponents for window length $w=250$ kyr computed from model run with $\tau = 1.45$ and $u = 0.15$: $X(t)$ (dotted blue), FTLE$([t-w,t])$ (gold).}
\label{fig:FTLE}
\end{figure}

Fig. \ref{fig:FTLE} shows the time profile of the largest FTLE along the trajectory.  Before the transition around 800 kyr BP, the FTLE generally remains negative apart from a few short excusions above zero.  At 1000 kyr BP the FTLE approaches zero and remains there for some time.  Just before 800 kyr BP it goes positive and then on average stays positive for the remainder of the trajectory.

The negative FTLEs leading up to the transition confirm that the trajectory forgets its initial history and the effect of past disturbances.  In particular, this implies that the
infinite-dimensional nature of the DDE's possible initial conditions
does not play a role for the MPT.  Whenever the system is exhibiting the small-amplitude response we observed negative FTLEs.

The positive FTLEs indicate a sensitivity of the trajectory during the large-amplitude response.  This sensitivity affects the
precise timing of the deglaciations, as we now demonstrate by
noise-induced desynchronisation of nearby trajectories.

To explore further the desynchronisation phenomenon outlined in \citet{desaed2013}, we add noise to our system.  The stochastic delay differential equation (SDDE) is then given as
\begin{eqnarray} \label{eq:noise}
dX(t) =&& [-pX(t-\tau)+rX(t)-sX(t-\tau)^{2} \nonumber \\
&&-X(t-\tau)^{2}X(t) - uM(t)]dt + \sigma dW(t),
\end{eqnarray}
Here, $W(t)$ is standard white noise and $\sigma$ is the noise amplitude.  We will always consider $\sigma$ as a fraction of the forcing amplitude $u$, i.e. $\sigma = \frac{u}{30}$.

\begin{figure}
\includegraphics[width=0.5\textwidth]{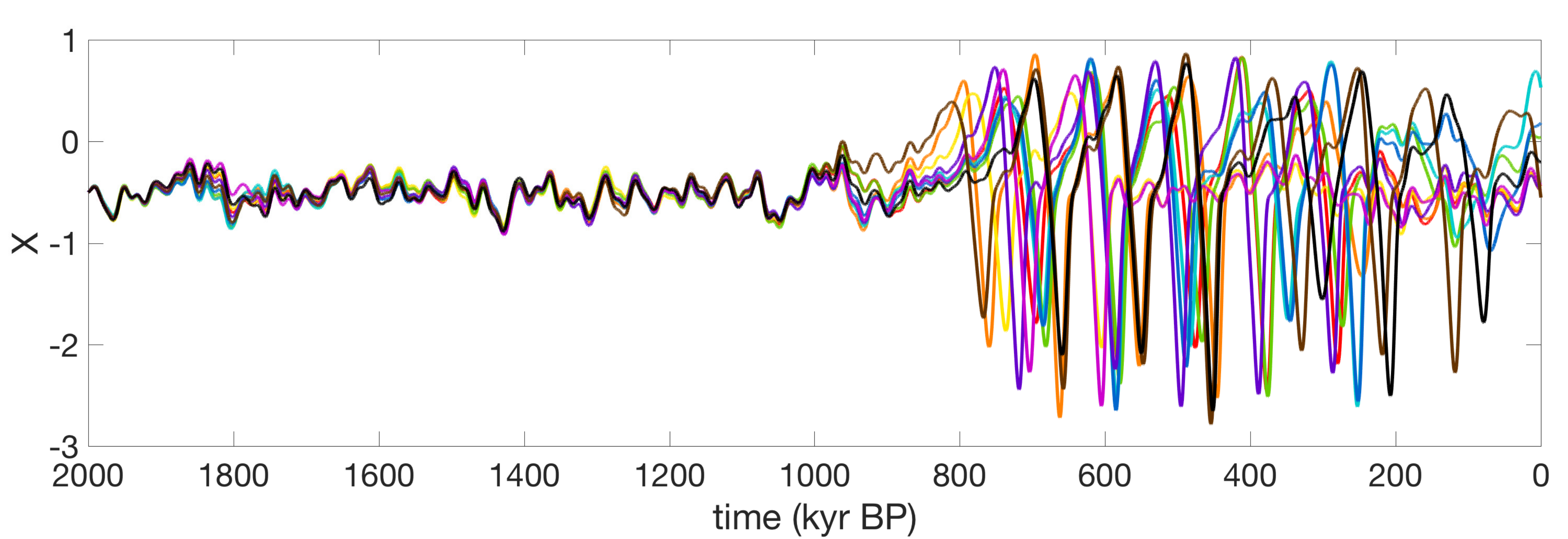}
\caption{Trajectories of DDE model with noise: $u=0.15$, $\sigma=0.005$, $\tau=1.45$.}
\label{fig:noise}
\end{figure}

\begin{figure}
\includegraphics[width=0.5\textwidth]{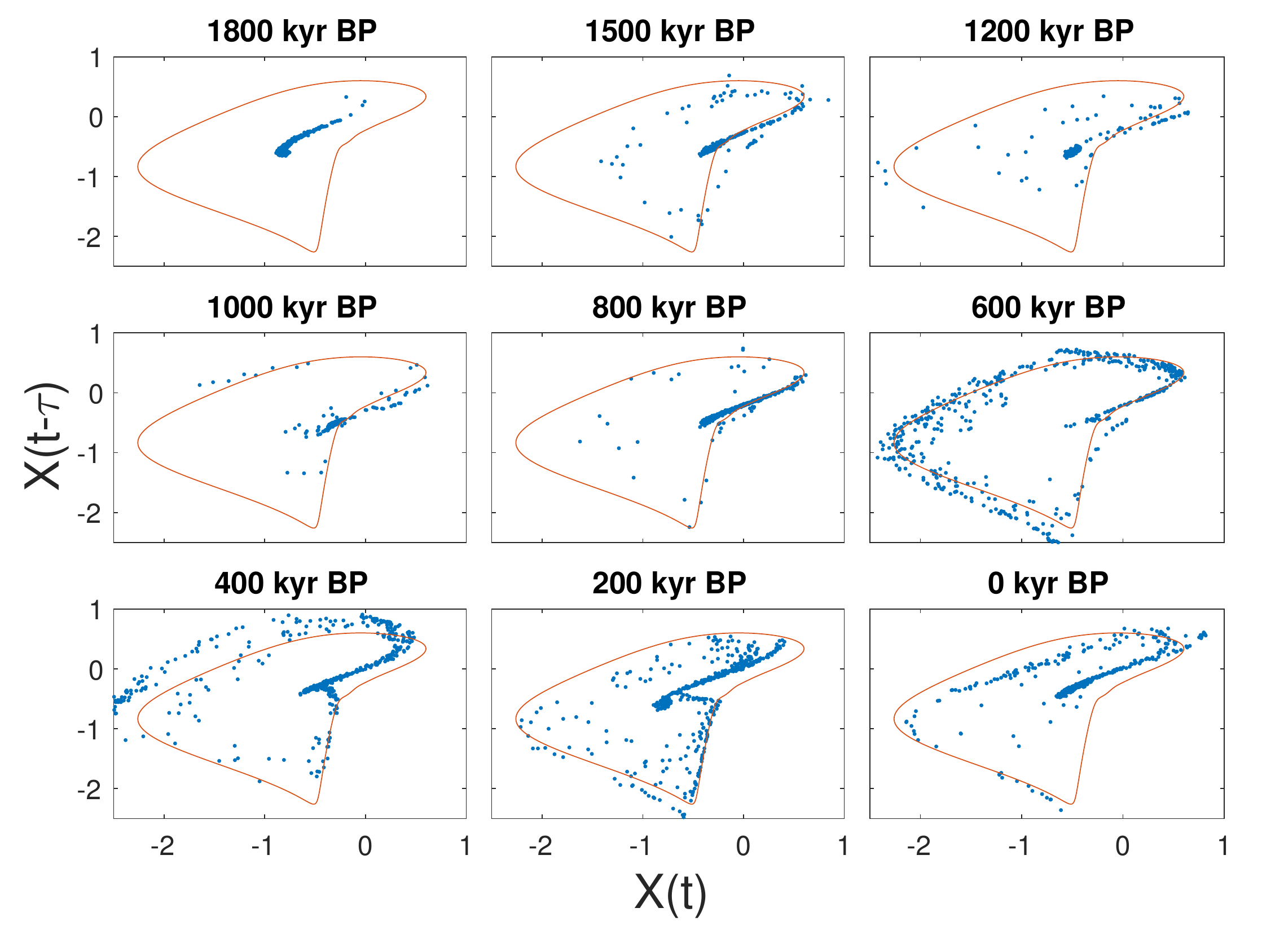}
\caption{Phase portraits at specific time instances of 500 model runs with noise: $u=0.15$, $\sigma=0.005$, $\tau=1.45$.  The red curve represents the unforced periodic orbit.}
\label{fig:phase_noise}
\end{figure}

\begin{figure}
\includegraphics[width=0.5\textwidth]{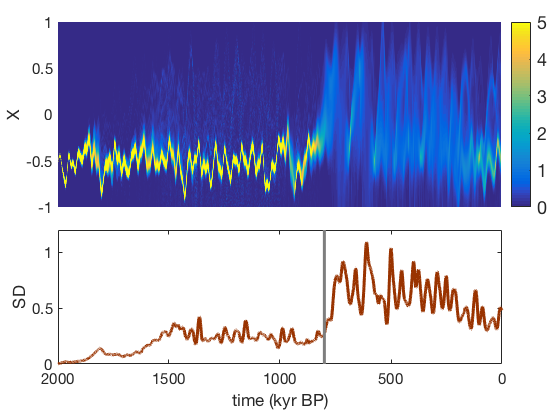}
\caption{Probability density in time (top) and its standard deviation (bottom) of 500 model runs with noise: $u=0.15$, $\sigma=0.005$, $\tau=1.45$.  The gray line indicates 800 kyr BP}
\label{fig:probdensity}
\end{figure}

We set the deterministic forcing strength $u=0.15$ and the noise
amplitude $\sigma = 0.005$.  We ran 500 realisations of the model, all
with the same initial history.  The results of 10 randomly selected
realisations can be seen in Fig. \ref{fig:noise}.  Up until 1000 kyr
BP all trajectories generally track the same solution with only minor
short desynchronisations.  Shortly after 1000 kyr BP the trajectories
begin to diverge, making the transition to the large amplitude
oscillation state at different times.  The trajectories then stay
desynchronised, which corresponds to being at a different phase along
the large amplitude oscillation.  We illustrate this phenomenon in
Fig. \ref{fig:phase_noise}, where we show the distribution of
trajectories in the $(X(t),X(t-\tau))$-plane along the unforced
periodic orbit, and in Fig.~\ref{fig:probdensity}.
Figure~\ref{fig:probdensity} shows the transition from
  synchronization to desynchronization in a density plot.  At approximately 800 kyr BP (indicated by the gray line) the probability density makes a sharp transition from a small-variance to a large-variance-low-maximum density.  The timing of this transition agrees with the first large positive excursion of the FTLE in the deterministic case and with the MPT.  The standard deviation shows this transition as well, but with a lag of about 50 kyr.
Figures \ref{fig:phase_noise} and \ref{fig:probdensity} also show that the distribution by desynchronisation is far from uniform. There are still distinct deglaciation times favoured for many realisations. For this reason the desynchronisation observed in figures \ref{fig:phase_noise} and \ref{fig:probdensity} does not contradict studies that have argued for phase-locking to different components of the orbital forcing.  One of the original hypotheses of phase-locking in the late Pleistocene was suggested to be related to eccentricity, specifically associated to events of low eccentricity \citep{hays1976,paillard2015}.  More recent studies have looked at the possibility of locking to precession or obliquity.  In \citet{ridgwell1999} the authors argue locking to every 4th or 5th precession cycle, while \citet{huybers2005} argue for locking to every 2nd or 3rd obliquity cycle.  Later, \citet{huybers2011} attributes locking to a combination of precession and obliquity, and states strongly that the pacing cannot be attributed solely to one of these components. Robust evidence for locking requires testing on a range of initial conditions or noise realizations (or long time series), but the short data record corresponds to only one realization of noise disturbance and one initial condition. Thus, data available may be insufficient to
  distinguish the higher-order locking proposed in the literature from the level of desynchronization we report in figures \ref{fig:phase_noise} and \ref{fig:probdensity}.

\subsection{Effect of noise with stronger forcing}

\begin{figure}
\includegraphics[width=0.5\textwidth]{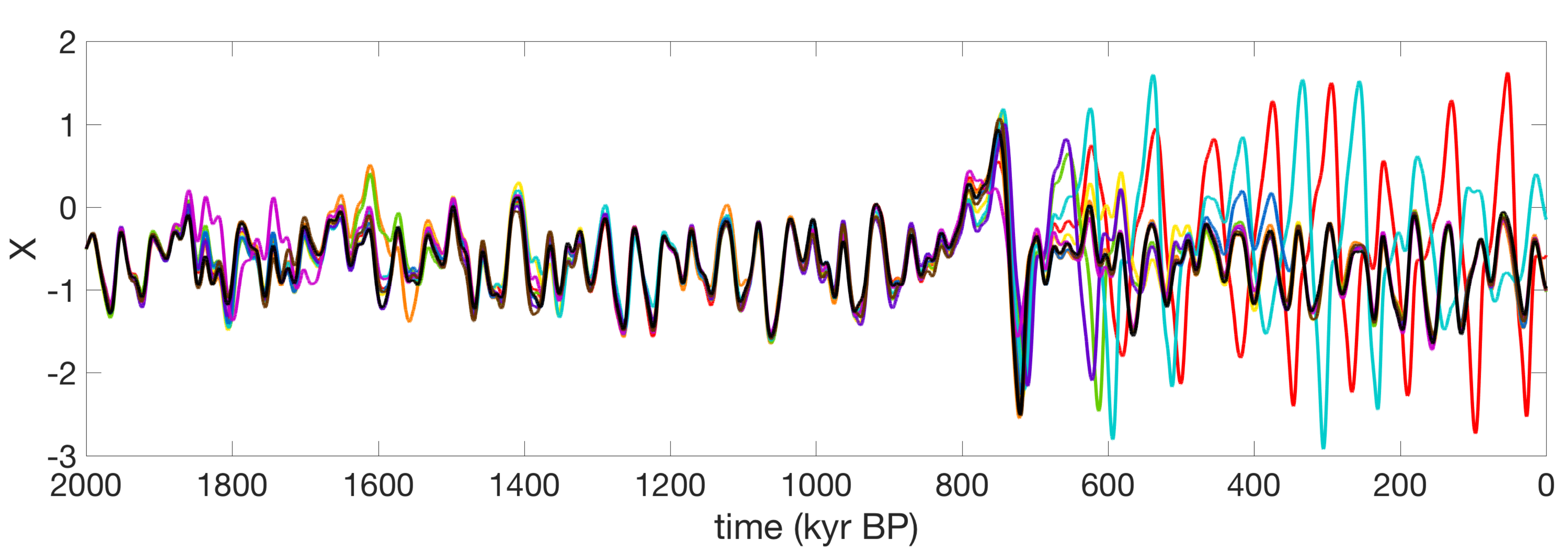}
\caption{Trajectories of DDE model with noise: $u=0.45$, $\sigma=0.015$, $\tau=1.45$.}
\label{fig:noise_u45}
\end{figure}

\begin{figure}
\includegraphics[width=0.4\textwidth]{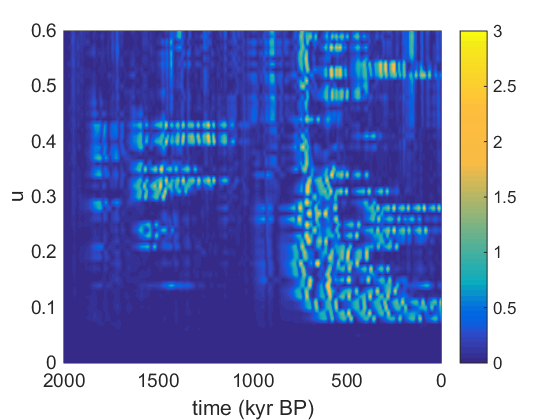}
\caption{Distance from reference solution at $\tau_{\mathrm{ref}} = 1.25$, for different forcing strengths $u$ and noise strength $\sigma = u/30$. Averages taken over window length of size $\tau = 1.45$. Initial condition $X(t) = −0.5$ for $t \in [2 + \tau , 2]$ Myr BP.}
\label{fig:disteq_tau145_noise}
\end{figure}

Figure \ref{fig:noise_u45} shows that the addition of noise enhances the MPT for strong forcing.  When adding noise, we observe transitions for forcing strengths $u$ for which there was no transition in the deterministic case (compare Figures
  \ref{fig:disteq_tau145} and \ref{fig:disteq_tau145_noise}).  In Fig.~\ref{fig:disteq_tau145} the last persistent transition occurs at $u=0.33$.  Fixing the forcing strength at $u=0.45$, we compute ten realisations.  Here, two realisations exhibit transitions that persist until the end of the run.  Although the transitions don't appear as commonly as in the weaker forcing case, with noise it is possible to observe an MPT-like transition across a wider range of forcing amplitudes.

Figure~\ref{fig:disteq_tau145_noise} shows a systematic overview
  of the transition enhancing effect of noise.  We added noise of amplitude $\sigma = \frac{u}{30}$ and compute the distance diagram as in Fig~\ref{fig:disteq_tau145}.  We observe transitions occuring above the maximal value of $u$ for which transitions occured in the
deterministic case.

\section{Discussion}

We have presented an analysis of the well known \citet{saltzman1988} model of Quaternary climate dynamics.  We have shown that it can be reduced to a scalar equation for the ice mass anomaly $X(t)$, by collecting all other variables into a delayed feedback term.  Through this formulation we are able to explore the dependence on the time scale $\tau$ of the delay in this feedback.  We discovered a region in which a stable equilibrium and stable large-amplitude periodic orbit coexist for the unforced system.  This parameter region is more consistent with physically relevant timescales for the ice sheet and deep ocean response than those explored in SM88.

We observe two different responses to astronomical forcing of the
type propsed by \citet{huybers2006}: small-amplitude response
(tracking the equilibrium) and large-amplitude response (tracking
the periodic orbit)  In the deterministic bistable region the model exhibits a switching behaviour between these two responses.  The timings of the switches are robust for different values of the delay and focing strength.  The most prominent switch occurs around 800 kyr BP which is within the range of when the Mid-Pleistocene Transition is believed to have occurred.

One trajectory showing the MPT is analysed using finite-time Lyapunov exponents.  Using a window of 250 kyr we observe that the largest FTLE is negative up to 1000 kyr BP, remains around zero during the transition, and is mostly positive after the transition.  We demonstrate that this positivity induces phase desynchronisation which is then confirmed by adding noise to the model.  The added noise also allows for the transition to be seen with stronger forcing. 

The novelty of this study is in the observation that transitions similar to the MPT occur robustly without any additional climate forcing, such as a nonstationary background state.  The transition arises from the bistability rather than a shift of a system parameter through a bifurcation.  We also highlight that with noise the transition is possible but doesn't always occur, and that the phase (or timings of the deglaciations) are variable (however still influenced by the forcing frequencies).  While our demonstration was done for the specific low-order model SM88, we conjecture many of the effects (instability caused by delayed feedback, transition induced but not caused by the forcing) to be present in other models with similar bistability.

Many models assume a slowly varying background state, typically related to slowly declining CO$_2$ levels, and attribute this as the primary mechanism for the MPT \citep{saltzman1991,tziperman2003,ashwin2015,tzedakis2017,paillard2017}.  Our proposed mechanism does not exclude the possibility of this changing background state.  However, it does not require the slow change of background state to cause a bifurcation. Instead, we consider a new null hypothesis for the MPT which involves only bistability and external (orbital and/or stochastic) forcing.  Future work could include an additional varying background state within the parameter regions mapped in Fig.~\ref{fig:bifdiag_dde} and analyse the effects on the transition.

We reported results for one specific astronomical forcing time series (65$^{\circ}$N summer integrated insolation), which is a weighted average version of the typical insolation threshold forcing (see Appendix~\ref{app:forcing}).  Qualitatively identical results can also be found when choosing a particular insolation threshold. However the particular threshold at which the behaviour is observed changes for different delays.  The weighted average over thresholds from Appendix~\ref{app:forcing} does not depend on an additional parameter (forcing threshold) and the dynamical behaviour is preserved.

Some questions that arise from this study will need to be addessed
by deeper mathematical analysis.  Why does the model prefer certain
times to switch, and what causes the sharp threshold in forcing
strength where switching becomes possible, visible in Fig.~\ref{fig:heatmaps}?  The latter can be answered by considering simple harmonic forcing with the dominant period of
 $41$ kyr seen in the astronomical forcing.  This threshold should be analysed further in a follow-up study.  Another question to consider is the effect of human activities on the system.  Is it possible anthropogenic contributions could knock the system out of the oscillations again, and perhaps to a completely different dynamical region?  Addressing this question would be useful for understanding current and future climate scenarios.

\appendix

\section{\label{app:LCA}The linear chain approximation}

A linear chain of first-order ODEs approximates a delay in the
following sense (see \citet{smith2010} for a precise derivation).  Here we adapt the derivation to our particular system.
The 3-dimensional SM88 model \eqref{eq:MS88}, using \eqref{eq:MS88-LCA}, has a 2-dimensional linear chain and is of the following form:
\begin{subequations} \label{eq:LCA_2D}
\begin{align}
\frac{\d Y}{\d t} &=F(Y(t),Z(t)), \\
\frac{\d V}{\d t} &= (Y(t) - V(t)), \\
\frac{\d Z}{\d t} &= q(V(t) - Z(t)).
\end{align}
\end{subequations} 
Extracting the linear chain system we have
\begin{equation}
\frac{\d\vec{y}}{dt} = A\vec{y}(t)+\vec{b}(t), \label{eq:system_a}
\end{equation}
where $\vec{y}=[V,Z]^T$.  We have $A$ and $\vec{b}(t)$ as follows:
\begin{equation}
A =\begin{bmatrix}
    -1  &  \phantom{-}0 		\\
    \phantom{-}q       	&  -q
\end{bmatrix},
\qquad
\vec{b}(t) =\begin{bmatrix}
    y_1(t)  \\
    0 
\end{bmatrix}.
\end{equation}
Since $A$ is invertible, the fundamental solution matrix of \eqref{eq:system_a} is then given by
\begin{equation}
\Phi(t) = P\e^{\Lambda t}P^{-1} \label{eq:fundsol_ai}
\end{equation}
where $P$ is the matrix of eigenvectors associated with eigenmatrix $\Lambda$,
\begin{equation}
\Lambda = \begin{bmatrix}
-1 & \phantom{-}0  \\
\phantom{-}0 & -q 
\end{bmatrix},
\end{equation}We have the following basis of $A$ and it's inverse,
\begin{equation}
P =\begin{bmatrix}
    \frac{q-1}{q}  &  0 		\\
    1       	&  1
\end{bmatrix},
\qquad
P^{-1} =\begin{bmatrix}
     \frac{q}{q-1} & 0  \\
     \frac{-q}{q-1} & 1 
\end{bmatrix}.
\end{equation}
From \cite{perko2013} we have the fundamental solution
\begin{equation}
\vec{y}(t) = \Phi(t)\vec{y}(0) + \int_0^t \Phi(t)\Phi^{-1}(s)\vec{b}(s) \d s.
\end{equation}
Assuming the solution has existed arbitrarily far in the past and using a change of variables we can extract an equation for $Z(t)$,
\begin{equation} \label{eq:z_sol}
Z(t) = \int_{0}^{\infty} Y(t-\tau)K(\tau)\d\tau\mbox{,}
\end{equation}
where the kernel in \eqref{eq:z_sol} is
\begin{equation}
K(\tau) = \frac{q}{q-1}\left[\e^{-\tau} -\e^{-q\tau}\right].
\end{equation}
Approximating the true distributed delay kernel $K$ in
\eqref{eq:z_sol} with a delta distribution at its expected value
\begin{align}
\mathbf{E}(K(\tau))=&\int_{0}^{\infty} \frac{q}{q-1}\left[\e^{-\tau} -\e^{-q\tau}\right]\tau \d\tau=\frac{q+1}{q}
\end{align}
we approximate \eqref{eq:LCA_2D} by a DDE with a discrete delay 
\begin{equation}
\frac{dY}{dt} =F\Big(Y(t),Y\Big(t-\frac{q+1}{q}\Big)\Big).
\end{equation}

\section{\label{app:forcing}Extraction of integrated summer insolation forcing $M(t)$ from data}

We use the Integrated Summer Insolation Calculations data set provided by \cite{huybers2006} (found at \url{https://www1.ncdc.noaa.gov/pub/data/paleo/climate_forcing/
orbital_variations/huybers2006insolation/j_65north.txt}), last accessed: 17 October 2018. The data is provided as daily average summer energy in GJ/m$^2$ calculated from number of days the insolation was above a given threshold (W/m$^2$).  We used the data provided for the latitude of 65$^{\circ}$ North.  Rather than selecting a particular threshold, we took the average of all the thresholds.  This has the effect of giving a linear weighting to the threshold reached in a day: days that reached higher insolation levels are given a proportionally larger weight. The precise expression is as follows,
\begin{equation}
I_{\mathrm{agg}}(t) = \sum_{i=1}^{25}I_{i}(t).
\end{equation}
Here $I_{i}(t)$ is the average daily summer insolation calculated by \cite{huybers2006} for the threshold $25(i-1)$ in year $t$ (column number $i+1$ in the data file).  We then normalise this data through
\begin{equation}
M(t) = \frac{I_{\mathrm{agg}}(t)-\operatorname{mean}_t I_{\mathrm{agg}}}{\operatorname{std}_t I_{\mathrm{agg}}},
\end{equation}
where $\operatorname{mean}_t I_{\mathrm{agg}}$ and
$\operatorname{std}_t I_{\mathrm{agg}}$ are the mean and standard
deviation of $I_{\mathrm{agg}}(t)$ over all times $t$.

\section{\label{app:FTLE}Computation of Finite-Time Lyapunov Exponents (FTLEs)}

Our algorithm for computing Lyapunov exponents for a delay differential equation (DDE) follows from \citet{farmer1982}.

\paragraph{Linearisation}

We consider a trajectory $X(t)$ of \eqref{eq:forced} and
linearize \eqref{eq:forced} along this trajectory:
\begin{equation}\label{eq:lindde}
  \dot x(t)=J_0(t)x(t)+J_\tau(t)x(t-\tau)\mbox{,}
\end{equation}
where $J_0$ an $J_\tau$ are the derivatives of the right-hand side of
\eqref{eq:forced} with respect to $X(t)$ and $X(t-\tau)$:
\begin{eqnarray}
J_{0}(t) &&=  r-X(t-\tau)^{2}, \\
J_{\tau}(t) &&=  -p-2sX(t-\tau)-2X(t)X(t-\tau).
\end{eqnarray}

\paragraph{Discretisation}

We consider an approximate solution of \eqref{eq:lindde} $x_i=x(t_i)$ for $t_i=t_0+ih$, where $h$ is small step size, obtained using a second order trapezoidal numberical solver for DDEs.  We then discretize \eqref{eq:lindde}  into $m = \tau/h$ steps,
\begin{equation}
\vec{y}_{i+1} = M_i  \vec{y}_{i}; \quad
\vec{y}_{i} = [x_{i},...,x_{i-m}]^{T},
\end{equation}
where $M_i$ is a square $(m+1)\times (m+1)$ matrix of the form
\begin{equation*}
M_i =\begin{bmatrix}
    M_{i,1}  &  0 		& \dots   	 & M_{i,m} & M_{i,m+1} \\
    1       	&  0 		& \dots    	& 0 	    & 0 \\
    0		& \ddots        &   	& \vdots      & \vdots \\
     \vdots      &         	& \ddots  	& 0      & \vdots \\
    0		 & \dots      	&  0      	& 1		& 0
\end{bmatrix}
\end{equation*}
with  entries
\begin{align*}
M_{i,1} =& 1+ \frac{h}{2}(J_{0}(t_{i+1})+J_{0}(t_i))+
\frac{h^{2}}{2}J_{0}(t_{i+1})J_{0}(t_i), \\
M_{i,m} =& \frac{h}{2}J_{\tau}(t_{i+1}), \\
M_{i,m+1} =& \frac{h}{2}J_{\tau}(t_i)+
\frac{h^{2}}{2}J_{0}(t_{i+1})J_{\tau}(t_i).
\end{align*}

\paragraph{QR method}

We use a continuous QR algorithm for computing the Lyapunov exponents (see \citet{dieci1997}).  The QR algorithm is based on the numerical linear algebra factorization of a matrix $M$ into an orthogonal matrix $Q$ and an upper triangular matrix $R$.  Although DDEs have an infinite number of Lyapunov exponents, this method allows us to compute up to $m+1$ through the discretisation.  We can choose to compute only the largest $l$ Lyapunov exponents by using a non-square $Q$.
We initialise an arbitrary othogonal $Q_0$ as an $m\times l$ matrix of form 
\begin{equation*}
Q_0 =\begin{bmatrix}
    I_{l} \\
	0_{(m-l) \times l} \\
\end{bmatrix}.
\end{equation*}
Note that $I_{l}$ is the $l\times l$ identity matrix and $0_{(m-l)\times l}$ is an $(m-l)\times l$ zero matrix.  We define interatively $Q_i$ and $R_i$ by the QR decomposition of $M_iQ_{i-1}$ (matlab command qr($M_iQ_{i-1},0$)),
\begin{equation}
Q_iR_i = M_iQ_{i-1},
\end{equation}
which produces a square $l\times l$ upper triangular matrix $R_i$ with eigenvalues $R_{i,jj}>0$ ($j = 1,...,l$).  We store $R_i$ for each timestep.  After $N$ timesteps we have the relation
\begin{equation}
Q_NR_N...R_1 = M_N...M_1Q_0.
\end{equation}
The infinite Lyapunov exponents can then be approximated by
\begin{equation}
\lambda_j = \frac{1}{N}\sum_{i=0}^{N}\ln R_{i,jj}.
\end{equation} 
For FTLEs we truncate the above sum after $W$ timesteps, and view how this truncation changes in time, i.e.
\begin{equation}
\lambda_{j,n} = \frac{1}{hW}\sum_{i=n-W}^{n}\ln R_{i,jj}.
\end{equation} 
Here $W = w/h$, where $w$ is the desired time window length.  Note that $n$ initialises at time $W$.

\begin{acknowledgments}
C.Q., J.S and T.L. have received funding from the European Union's
Horizon 2020 research and innovation programme under the Marie
Sklodowska-Curie grant agreement No 643073. J.S. gratefully
acknowledges the financial support of the EPSRC via grants
EP/N023544/1 and EP/N014391/1.
\end{acknowledgments}

\bibliography{references}

\end{document}